\documentclass[11pt,hyphens]{article}
\usepackage{amsmath}
\usepackage{amssymb}
\usepackage{bm} 
\usepackage[usenames]{color}

\usepackage{fancyhdr}
\usepackage[T1]{fontenc}
\usepackage{graphics}
\usepackage{graphicx}
\usepackage[]{hyperref}
\hypersetup{backref=true, 
        citebordercolor=1 1 1, 
        linkbordercolor=1 1 1,
        urlbordercolor=1 1 1,
        pdfborder= 0 0 .5, 
        }  
\usepackage[latin1]{inputenc}
\usepackage{lscape}
\usepackage{lastpage}
\usepackage[longnamesfirst]{natbib}
\usepackage{setspace}
\usepackage[showlabels, sections, floats, textmath, displaymath]{preview}
\usepackage{psfrag} 

\usepackage{rotating,epsfig} 
\usepackage{shadow}
\usepackage{shapepar}
\usepackage{subfigure} 

\usepackage{verbatim}

\usepackage{times}

\newcommand{\noi}{\noindent}

\bibpunct[;]{(}{)}{,}{a}{,}{;}

\newcommand{\monthname}[1]{%
\ifcase#1
\or January%
\or February%
\or March%
\or April%
\or May%
\or June%
\or July%
\or August%
\or September%
\or October%
\or November%
\or December%
\fi }

\input{pst-plot}

\graphicspath{{/Users/sahap/PoolForPrivacy/Codes/}}

\title{
  Covariate Microaggregation for Logistic Regression: An Application for
  Analysis of Confidential Data
}
\author{Paramita Saha-Chaudhuri\\ \texttt{paramita.sahachaudhuri@mcgill.ca}}
\date{
Updated: \today}
\begin{document}
\topmargin -0.5in
\maketitle
\color{black}
\vspace{-10mm}

\abstract
\noi
In the recent past, electronic health records and distributed data
networks emerged as a viable resource for medical and scientific
research. As the use of confidential patient information from such
sources become more common, maintaining privacy of patients is of
utmost importance.  For a binary disease outcome of interest, we show
that the techniques of microaggregation (equivalent to specimen
pooling) and \underline{Po}oled 
\underline{Lo}gistic \underline{R}egression (PoLoR)
could be applied for analysis of large and/or distributed data while
respecting patient privacy. PoLoR is exactly the same as standard
logistic regression, but instead of using individual covariate level,
the analysis uses microaggregated covariate level when
microaggregation is conditional on the outcome status. Aggregate
levels of covariates can be passed from the nodes of the network to
the analysis center without revealing individual-level microdata and
can be used very easily with standard softwares for
estimation of disease odds ratio associated with a set of categorical
or continuous covariates.  Microaggregation of covariates 
allows for consistent estimation of the parameters of logistic
regression model that can include confounders and transformation of
exposure. Additionally, since the microdata can be 
accessed within nodes, effect modifiers can be accommodated and
consistently estimated. For analysis of confidential health data,
covariate microaggregation for logistic regression will provide a
practical and straightforward alternative to more complicated existing
options. 

\section*{Keywords}
Data privacy; Distributed data; PoLoR (\underline{Po}oled
\underline{Lo}gistic \underline{R}egression).

\section{Introduction}

Each year in the US and elsewhere, state and federal governments
and health care authorities collect immense amount of personal and
sensitive data on health care use, diagnoses, risk factors and
behaviors and many others features. The advent of personalized
medicine and the genomic revolution will result in further
gathering of a large number of very sensitive data including genomic
data on large segments of the population. Although these data could
provide critical information for the management of the health care
system, for studying causation of diseases, prognosis and the impact
of treatment or prevention efforts, their use is constrained by
legitimate concerns about privacy of information. 
Concerns about data privacy is not just limited to healthcare field.
The issue has long plagued statistical agencies throughout the
world. As a result, many agencies are bound by law to 
protect the information they collect from individuals and 
businesses. For example, US Census Bureau is bound by Title 13 of the
US Code. In order to extract useful information from sensitive
individual-level data (microdata), agencies frequently employ various
Statistical Disclosure Limitation (SDL) (also known as, Statistical
Disclosure Control (SDC)) techniques that strive to achieve a
balance between data use and confidentiality with varying degree of
complexity and success. 

With the rise of data networks for pharmacosurveillance, such as 
the Sentinel Initiative of the USA Food and Drug Administration (FDA)
\citep{CookEtAl2012MiniSentinelCausalRiskDiff} 
and Canadian Network for Observational Drug Effect Studies (CNODES)
\citep{CNODES}, data confidentiality is now paramount in healthcare
and clinical research. An additional 
complexity with such large data network is that the data is
distributed over several nodes and frequently, law limits sharing of
the data between nodes and even with the analytical center (AC)
\citep{HIPAA,  Ness2010AJEPrivacy}. Therefore, even though ideal 
\citep{NattingerEtAl2010AJEPrivacyHonestBroker}, it is not
possible to combine and create a single analysis dataset. Statistical
analysis and release of summary statistics, such as tables of
population characteristics, parameter estimates, etc. need to comply
with privacy laws that severely restrict analytical toolset. For
example, for exploratory analysis of data and creating tables, most
provincial partners of CNODES restrict release of table shells with
less than $5$ observations. Exploration of association between
variables is even more complex.

A primary goal of pharmacosurveillance using these large health
databases, such as CNODES, is to identify post-marketing safety signal
of drugs 
\citep{YuEtAl2014IncretinCHF}. Existing approaches for modeling an
adverse effect of a drug can be classified as either non-interactive
or interactive \citep{Dwork2006DiffPrivate}. In the non-interactive
approach, each node or data owner performs analysis on their own
dataset and the results are combined using meta-analytic techniques
\citep{Filion2014GutPPI, Wolfson2010IJEDatashield}.  A drawback of
this approach is that exploratory analysis or model selection may be
cumbersome or may not even be possible. Interactive approach is more
involved and requires nodes to 
work together to pass summary statistics to and from AC
\citep{Wu2012JAMIAGLORE, ElEmamEtA2013JAMIASecureLogistic}. For a
continuous outcome and linear regression, analysis is relatively
straight-forward using either approach, but privacy-preserving
modeling options of a binary disease outcome is limited and the
existing approaches are quite complicated
\citep{FienbergEtAlSecureLogLinLogitReg,
  FienbergEtAl2009LogitRegMultupleSources} putting significant
analytical burden on potentially ill-equipped nodes with oversight
from the AC. While some of these approaches could be
adapted, the performance of the strategies have not been explored for
other types of designs, such as a
matched case-control design.


Building on our previous work on specimen pooling, in this manuscript,
we propose covariate microaggregation to ensure 
data confidentiality with the goal of estimating parameters of a
logistic regression model for a binary disease
outcome. We call this approach PoLoR (\underline{Po}oled
\underline{Lo}gistic \underline{R}egression). PoLoR retains the flavor
of interactive 
approach with a ease of computation similar to the non-interactive
approach. For estimating covariate odds ratio (OR), rather than
sharing the summary statistics at each stage of iteration, we propose
to share aggregate covariate information \underline{only once} from
the nodes to the AC. PoLoR does not require multiple
updates for parameter estimation because the aggregate covariate
information can be used in a single pass to estimate relevant model
parameters. Since only aggregate covariate 
information is exchanged, there is limited concern about privacy. The
approach requires individual nodes to only aggregate or add
covariates according to a predefined protocol and does not impose any
additional computational or statistical burden. Moreover, since the
approach is equivalent to the standard logistic regression, but on
aggregate covariate level, standard error estimates of the model
parameters can be obtained and model selection using likelihood ratio
test is also possible and valid. 

Microaggregation of continuous variables has a long history in SDL
literature   \citep{Domingo-FerrerMateo-Sanz2002Microaggregation}.    
The connection between microaggregation and specimen pooling
\citep{WeinbergUmbach1999Biometrics, SahaPooledMatching2011,
  SahaPooledTTP2010} has not been recognized before, but both, in
principle are the same. To obtain microaggregates for an
individual-level data or microdata, data from multiple participants
are aggregated to produce the average of the values for the
group. While microdata is not disclosed to outsiders (e.g., AC),
insiders (a particular node) need to 
have access to the microdata in order to perform aggregation.
The context for specimen pooling is different than
microaggregation. When environmental and/or chemical exposures are to
be measured in biospecimen, e.g., urine, using an expensive assay,
specimen pooling offers an efficient strategy that provides savings for
both monetary resource as well as (banked)
biospecimen. Here, individual-level exposure is not known in
advance. Instead, biospecimen from multiple participants are
aggregated or combined according to a rule to produce a pooled
specimen. The pooled specimen is then assayed to measure the
exposure. Consequently, only the average exposure for the group is
available and individual-level exposure is not available at all. For a
binary 
disease outcome, pooled exposure can be used to estimate
individual-level exposure-disease 
OR \citep{WeinbergUmbach1999Biometrics, SahaPooledMatching2011,
  SahaPooledTTP2010} and the strategy is most useful when outcome
ascertainment is inexpensive, but exposure assessment is expensive.


Leaning on the connection between microaggregation and specimen
pooling, we demonstrate that the aggregate covariates can 
be used to estimate individual-level OR for a covariate with a binary
outcome. 
The rest of the manuscript is organized as follows. In section
\ref{sec:methods}, we outline the PoLoR  approach for a binary outcome
and microaggregated covariate.
We then demonstrate how PoLoR can
be applied for confidential data while protecting patient privacy. 
In section \ref{sec:example}, we demonstrate the application of
PoLoR with simulated and real 
datasets and compare the results
with standard logistic regression analysis. We make some practical
recommendations and conclude with a discussion.
\section{Methods}
\label{sec:methods}

In this section we outline the covariate microaggregation strategy and
describe PoLoR when it 
is of interest to estimate the ORs associated with a set of covariate
for a binary outcome (e.g., presence and absence of a disease). We
then extend the concept for application with 
distributed data network. Since the analysis is based on
aggregate rather than microdata, even when the data pass through
firewalls, patients' 
identity are always protected. We demonstrate the approach for a
binary outcome. Categorical outcomes with multiple category logit
model can be easily accommodated whereas the approach for a continuous
outcome is obvious and has been studied
\citep{Domingo-FerrerMateo-Sanz2002Microaggregation}.  

\subsection{Specimen Pooling}
PoLoR is based on the idea of pooling or combining biospecimen such as 
blood or serum. Instead of analyzing individual assay for exposure or
biomarker measurement, the idea is
to combine specimen from multiple assays and measure exposure in the
combined assay or pools. Thus, instead of knowing individual
level of exposure, only the average exposure
can be measured in the pools. In contrast, for microaggregation,
microdata is available to inform pooling strategy. But after the pools
are formed, only the average covariate level of the pools are reported
and/or shared. This connection between specimen pooling
and microaggregation has not been recognized before, but both strategy
are the same in principle, where instead of the individual level
covariate values, only group averages are recorded and shared.

The idea of specimen pooling originated during World War II in a
slightly different context when blood from military 
recruits were pooled to identify recruits with syphilis
\citep{Dorfman1943PoolingSyphilis}. Pooling 
of  biospecimen is used predominantly in infectious disease where
multiple specimens are pooled together to identify the infection
status of the samples, because if all of the contributing samples are
negative for an infectious disease (e.g., HIV), the pooled specimen
will also be negative whereas the pooled specimen will be positive
otherwise. Such a strategy leads to a reduction in expected assay
cost. Other uses of specimen pooling include application 
for estimating disease prevalence, to characterize distribution of
variables \citep{Caudill2010}, to assess diagnostic accuracy
\citep{FaraggiEtAl2003StatMed, LiuSchisterman2003PoolingROC},
etc.. DNA pooling has also been employed for association studies 
\citep{Sham2002DNAPooling}. The particular application of pooling that we focus
on is for estimating exposure OR for a logistic regression model for a
binary disease outcome \citep{WeinbergUmbach1999Biometrics,
  SahaPooledMatching2011} as we will discuss in detail later. 
We extended this approach for discrete survival time
outcome \citep{SahaPooledTTP2010}.  
In all these situations, pooling
offers an economical solution for various estimation problems.

\subsubsection{Notation}
We start with a simple scenario where we are interested to quantify
the effect of a covariate or exposure $X$ (continuous or categorical
with appropriate set of indicator variables)
on a binary outcome of interest $Y$ such as presence ($Y=1$) or
absence ($Y=0$) of disease. A logistic regression model can 
naturally be conceived for such a setting:
\begin{eqnarray*}
  \mbox{logit}(\Pr(Y=1|x)) = \beta_0 + \beta_1 x.
\end{eqnarray*}

To estimate the parameter of interest $\beta_1$, we can employ either
of the two designs. In a prospective study setting, we can randomly
sample subjects from a defined 
cohort and ascertain their exposure level and their outcome
status. This approach is not efficient for studying
rare diseases, so 
an alternate is to employ a case-control design where subjects are
selected based on their outcome status, followed by exposure
determination. While an unbiased estimate of the baseline log odds
$\beta_0$ can only be obtained from a prospective design, 
we can estimate $\beta_1$, the log OR associated with unit
increase in the level of $X$ from both prospective and case-control
design. In many scientific situations, we are primarily interested in
the exposure log OR $\beta_1$. While it is possible to estimate disease
prevalence from pooled data as is done in group testing, we assume
that the 
individual level outcome and baseline disease prevalence (and disease
odds) are known at the outset. For the rest of the manuscript, we use a
general notation keeping in mind that we are interested in estimating
exposure effect ($e^{\beta_1}$) rather than baseline disease odds
($e^{\beta_0}$). 

When microdata is available, $\beta_1$ can be estimated
easily. However, we are interested to  
characterize the association of $X$ and $Y$ {\it without using
  individual-level covariate information}, but using aggregate
covariate information only. Thus, to characterize the association of
exposure or covariate with a 
binary outcome via pooled approach, we require that the outcome
status of each subject be known in advance. This in turn implies
that the outcomes are easy and/or inexpensive to ascertain while the
exposure or marker of interest is expensive to assess. The proposed
pooling approach is applicable to a  case-control study as
well as a prospective or a cross-sectional study with a binary outcome  
\citep{WeinbergUmbach1999Biometrics, SahaPooledMatching2011}. 

Pooled analysis with a binary outcome has two stages: design stage and
estimation stage. In the design stage, we create pools or groups 
of observation to form the basis of aggregation and measure aggregate
covariate information for the pool. In the estimation stage,  we use
the pooled covariate measurements instead of individual covariate
measurements to estimate $\beta_1$ or covariate log OR.
Let $n$ denote the number of case subjects with $Y=1$ and $m$ denotes
the number of control subjects $Y=0$.  
Although pools of multiple sizes can be utilized for one study, for
simplicity of notation, we assume that pool of size $g$ is formed 
within cases and controls that both $n,m$ are multiples of $g$: $n
= k_n\times g$, $m = k_m \times g$, that is, each pools consists of
either $g$ cases or $g$ controls.

\subsubsection{Utility of Pooling in Epidemiologic Setting}
To see the utility of pooling in epidemiologic
settings, let us consider an example. Consider the association of
perfluorooctanoic acid (PFOA) and perfluorooctane 
sulfonate (PFOS) with adverse birth outcomes. In particular,
researchers have identified weak association of PFOA and PFOA with
preeclampsia \citep{Stein2009AJEPFOAPFOAPreeclamsia}. Preeclampsia is
a complication of pregnancy characterized primarily by high blood
pressure and proteinuria. The only treatment for  preeclampsia is
delivery of the baby. It is relatively easy to diagnose preeclampsia by
monitoring blood pressure and amount of protein in urine. However,
assessing the level of PFOA or PFOS in the blood costs at least
US\$400 per 
assay. Thus, a study recruiting only 50 preeclampsia cases and 50
controls without preeclampsia would require US\$40,000 simply to
analyze participants blood to assess the level of only one of the two
chemicals. If pooling in groups of two ($g=2$) is 
employed, then instead of analyzing 100 (50 from cases and 50 from
controls) specimens, we would only require to analyze 50 (25 case
pools and 25 control pools) specimens. Consequently, the total
exposure assessment cost will be cut down by half. 

Pooling also preserves valuable biospecimen. For example, suppose
assessing 
PFOS in blood requires 1ml of blood per assay. If, as before, $g=2$ is
employed, then we would need 0.5ml of blood per person so that the total
volume of the pooled specimen is still 1ml. If specimens from
participants are banked and/or cannot be recollected, then pooling
offers a viable strategy to save valuable biospecimen.

\subsubsection{Pooling Strategy and Covariate Assessment}

The first stage for pooled analysis is to create pools of observation
to form the basis of aggregation. Instead of measuring individual
level covariates, we measure aggregate or average covariate level for
the pool. So, irrespective of the number of subjects in a pool, we
obtain {\it one measurement per covariate for the pool}. Hence the unit 
of analysis is pool rather than individual participant.
To estimate the parameter of interest $\beta_1$, we create random
pools conditional on the outcome status. That is, for PoLoR, we create
pools of a given size by randomly pooling cases 
with cases and randomly pooling 
controls with controls. If we select a pool size of $g$, then to create
case pools (defined by $Y=1$), we randomly partition the cases in
$k_n$ groups each with $g$ subjects, such that each subject belong to
one and only one pool. Similarly, we partition $m$ controls (defined
by $Y=0$) to create $k_m$ groups of size $g$ each. For each group, we
simply evaluate the group-sum (or group average) of the covariate,
rather than reporting individual-level covariate values. PoLoR is
ultimately based on these pooled group-level or microaggregated
covariate values and does not involve individual-level covariate
information. 

\subsubsection{Pooled Model}

Suppose $Y_{ij}, j = 1, 2, \ldots, g$ denotes the outcomes of the
subjects belonging to the $i^{th}$ pool, $i=1, 2, \ldots, k_n$ for cases
and  $i=1, 2, \ldots, k_m$ for controls. Similarly, let $X_{ij}, j = 1,
2, \ldots, g$ denotes the corresponding exposure. 
For pooled analysis, we do not observed individual exposure 
measurements $X_{ij}, j = 1, 2, \ldots, g$, rather, for the $i^{th}$
pool, we only observe $\bar{X}_i = \frac{1}{g} S_i=\frac{1}{g}
\displaystyle\sum_{j=1}^gX_{ij}$, the pooled 
exposure level. It can be shown that
\citep{WeinbergUmbach1999Biometrics}:
\begin{eqnarray*}
  \mbox{logit}(\Pr(Y_{i1}=1, Y_{i2}=1, \ldots, Y_{ig}=1| \displaystyle
  s_i)) = \beta_0^*~g + \beta_1~s_i + \ln(r_{g})
\end{eqnarray*}
where $\beta_0^*= \beta_0 + \ln\left(\frac{\Pr(Y=0)}{\Pr(Y=1)}\right)$
and $r_g$ denotes the number of case sets of size $g$ divided by the
number of control sets of size $g$ in the data or $k_n/k_m$. Pooled
analysis also allows use of different pool sizes to accommodate for
the fact that $m$ and $n$ may not always have a common divisor. The only
restriction is that pools of a particular size should be represented
among both case pools and control pools. Thus, if pools of size 2 and
5 are used for cases, the same pool sizes should also be used for
controls and $r_g$ for each value of $g$ should be evaluated and used
in the model appropriately. 

Thus, a logistic regression based on the pools rather than the
individual outcomes, retain the same parameter of interest $\beta_1$
as the individual outcome model. Hence, the OR
for the covariate $X$ 
can be estimated using pooled measurements and outcomes. Moreover,
we do not need any novel tool or software to run a logistic
regression model with pooled covariates. Any statistical software capable
of estimating parameters of a logistic regression model can be used
with appropriate offset to estimate the parameters based on the
pooled model.

Confounders can be handled in a similar fashion by aggregating the
confounder measurements over the pools and including them in the
logistic model. Both categorical and continuous confounders can be
accommodated in this way.

For pooled analysis in a epidemiologic setting, which we call
traditional pooled analysis, only {\it categorical} effect 
modifiers are accommodated by pooling conditional on the levels of
{\it effect modifiers (EMs) and outcome status}. For example, if age (young
versus old) is an effect modifier, then pooling has to be done
separately within four strata defined as: (1) Young cases, (2) Old
cases, (3) Young controls and (4) Old controls.
However, pooling within the levels of effect modifier renders the main
effect of the effect modifier aliased with the baseline log odds and
hence it is not possible to estimate the main effect of the effect
modifier. Moreover, if multiple EMs are to be included in the model,
the advantage of pooling is reduced due to potential sparsity as
all combination of EMs and outcome may not have enough subjects ($g$
or more) required for pooling.

Another limitation of traditional pooled analysis is that
transformation of variables that are measured in pool cannot be
accommodated. Once a pool is formed, and aggregate level of the
covariate is measured in pool, individual level of the covariate
cannot be assessed without requiring a reanalysis of individual
specimen. This is because, in general for any non-linear function $h,
\sum_i h(x_i) \neq h(\sum_i x_i)$ and we can only measure $\sum_i x_i$ in
a pooled sample. Thus, a model such as:
$~~~\mbox{logit}(\Pr(Y=1|x)) = \beta_0 + \beta_1 \log(x),~~~$
cannot be employed in a pooled setting, because this model would
require evaluating $\sum_i \log(x_i)$ in the pooled sample whereas we
can only assess $\sum_i x_i$ in the pooled sample and cannot substitute
$\sum_i \log(x_i)$  with  $\log(\sum_i x_i)$.

\subsubsection{Model Diagnostics}
For a binary outcome and a logistic regression model, we can think of
two types of model misspecification: first, appropriateness of logit
link function, and second, selection between two competing logistic
regression model. The misspecification of the logit link is not
severe in general. Unfortunately, existing pooling methodology only
allows a logit link function for a binary outcome and hence does not
allow researchers to 
choose between different link functions and compare model fitting
between them. Hence, for model diagnostic with pooled covariate level,
we begin by assuming that the logit link function is appropriate for
the covariate-outcome association. As such, the second issue of
choosing between different logistic models is viable using pooled
data, especially in the context of distributed data setting. 

In a traditional setting, it is possible to assess association of the
exposure with the outcome using pooled exposure level. The pooled
model is simply a logistic regression model with pooled exposure level
instead of individual level exposure. Hence, in addition to the
parameter estimate, estimated standard error (SE), confidence interval
and associated 
p-value can be obtained for the exposure and other variables. If 
confounders are not subject to pooling, then alternate transformation
for confounder can also be evaluated. However, model selection
involving different transformations of the pooled variables will not
be possible. In addition,
we can also test for the interaction parameters for categorical EM
provided we pooled stratified by both outcome and levels of
EM. If two competing models can both be fit using pooled data, we can use
likelihood ratio test or AIC, etc.  ROC curve can be employed
to assess prediction accuracy of the model.
Noting that pooling fundamentally alters the intercept parameter akin
to a case-control study, model calibration tests such as
Hosmer-Lemeshow goodness of fit test cannot be employed
directly. If supplementary information about the disease prevalence is
available, then alternate approaches such as that of
\cite{HuangPepe2010SiMHosmerLemeshowForCaseControl} can be employed
for assessing model calibration. 

\subsection{Application of Specimen Pooling for Distributed Data} 
\label{subsec:application}
\label{subsubsec:distributed}
Coupling our approach with additional security measures such as secure
summation, we can directly apply  
pooled analysis strategy for a distributed data network. Whether the
data is horizontally or vertically partitioned, PoLoR can be adapted
to estimate the parameters of interest in a 
single pass, without requiring multiple iterations. We demonstrate the
implementation for a horizontally partitioned dataset where each node 
has exactly the same variables but holds only a subset of the
subjects; the implementation for a vertically partitioned dataset
where nodes hold different variables for a common set of subjects, is
similar. We consider the following scenario where there are multiple
nodes, each holding a subset of records and an analytical center that
can have contact with every node. A designated 
node can act as the center or representatives from nodes may
constitute a center or a center can be a suitable third party.
\begin{figure}[!htbp]
  \centering
  \includegraphics[scale=.5]{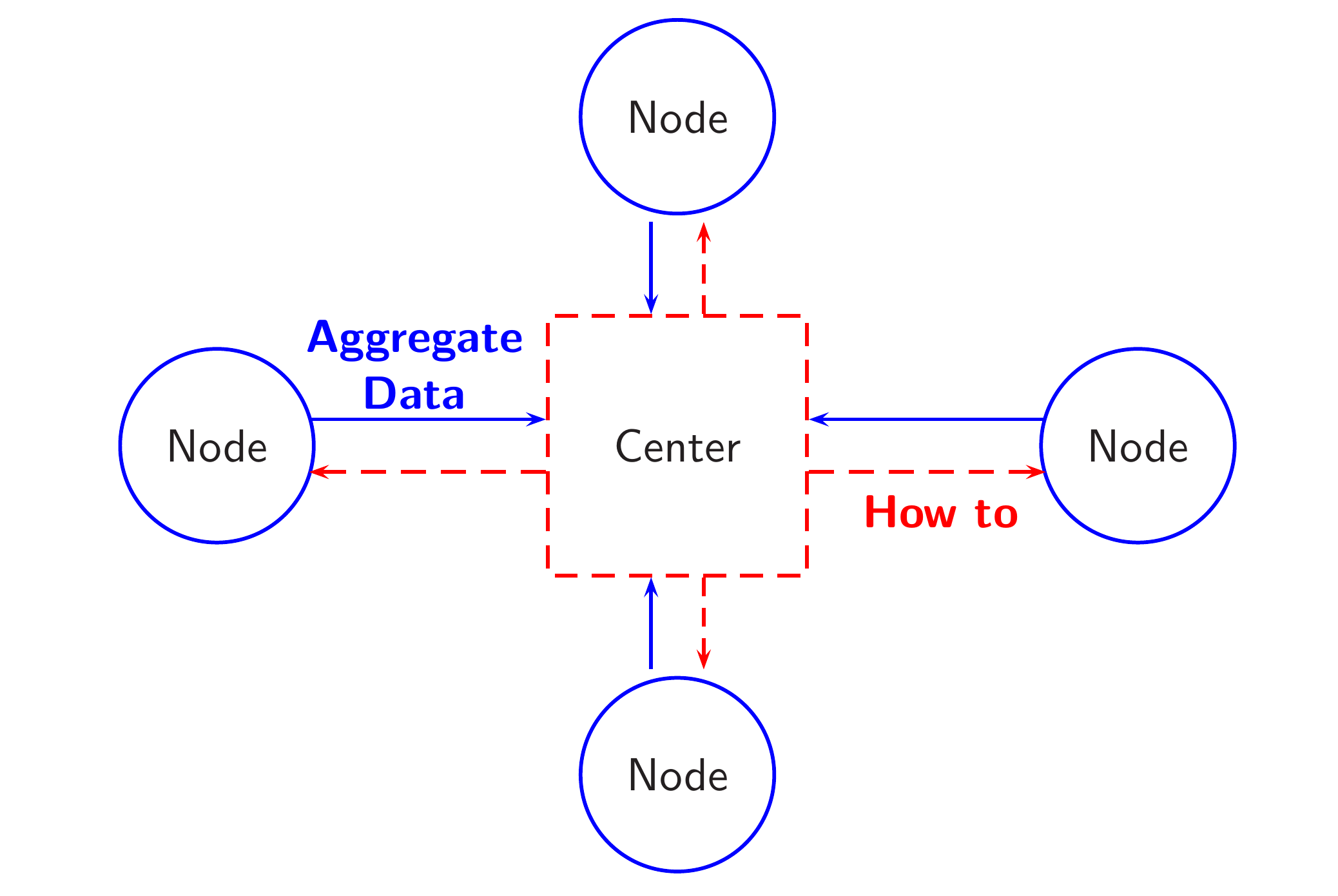}
     \caption{Schematic of a Distributed Data Network. Here we assume
      that the data is horizontally partitioned among the
      nodes. Center can send instruction to nodes as to how to
      aggregate data. Only aggregate data from the nodes can be passed
      to the center.} 
  \label{fig:DistributedData}
\end{figure}

\noi {\bf Implementation}\\
We suggest the following implementation of PoLoR
for the distributed data setting where patient privacy
need to be maintained throughout. By protecting patient privacy, we
mean that individual record, covariate level, outcome status, etc. of
a patient cannot be shared outside the specific node and we begin with
anonymized  data. We also assume that while individual
information cannot be shared between nodes or with the center,
aggregate level covariate information such as the total counts for
binary or categorical variables and sum/average for continuous
variables can be shared between nodes and centers. We assume that
agencies do not collude with each other and use secure summation in
addition to aggregate covariate information to further protect individual
record. We assume that the center can direct the nodes as to how to
aggregate the covariates and can receive aggregate data from nodes
(Figure \ref{fig:DistributedData}).

The implementation begins with the analytical center identifying the
number of case and control subjects in the study cohort. Each node can
send the number of cases and controls to the center without
compromising privacy. The next step is to determine the pool size $g$
and creating random partition separately among cases and controls to
identify which patients will contribute to which pool. To achieve this
with, say, $n = g \times k_n$ cases and 
$m = g \times k_m$ controls, we randomly partition the cases and
controls in $k_n$ and $k_m$ groups respectively. Suppose, subjects
$i_1, i_2, \ldots, i_g$ belongs to the $i^{th}$ case pool ($i = 1, 2,
\ldots, k_n$), and subjects $j_1, j_2, \ldots, j_g$ belongs to the
$j^{th}$ case pool ($j = 1, 2, \ldots, k_m$). These pool memberships
are passed onto the networks nodes to create pooled covariate
values. After the pooled covariate values are returned to the
analytical center, an analysis dataset is created based on the pooled
covariate values and logistic regression is run on the pooled values
to estimate covariate-specific ORs. 

The process can be described step by step as follows:
\begin{enumerate}
\item After applying appropriate inclusion and exclusion criteria,
  determine the number of cases ($n$) and controls ($m$) in the study
  cohort. 
\item Determine the pool size $g$ (if $n$ and $m$ have common divisor)
  or multiple pool sizes. 
\item Assuming $n = g \times k_n$, randomly partition cases in $k_n$
  groups with case ids $i_1, i_2, \ldots, i_g$ belonging to the
  $i^{th}$ case pool. 
\item Assuming $m = g \times k_m$, randomly partition controls in
  $k_m$ groups with control ids $j_1, j_2, \ldots, j_g$ belonging to the
  $j^{th}$ control pool. 
\item Pass a case-id dataset to the nodes of the networks (see Table
  \ref{tab:case-id-dataset}) to aggregate covariate for the cases.
\item Pass a control-id dataset to the nodes of data networks similar
  to the case-id dataset to aggregate covariates for the controls. 
  The case-id dataset and the control-id dataset will be the basis for
  pooling covariate  values.

\begin{table}
  \caption{Case-id dataset will consists of the pool id and the case
    ids contributing to specific pool.}
\label{tab:case-id-dataset}
\centering
    \begin{tabular}{lcccc}
      \hline \hline
      Pool id & case id 1 & case id 2 & $\ldots$ & case id $g$ \\ \hline
      $1$ & $1_1$ & $1_2$ & $\ldots$ & $1_g$ \\
      \vdots & \vdots & \vdots & $\ddots$ & \vdots \\
      $i$ & $i_1$ & $i_2$ & $\ldots$ & $i_g$ \\
      \vdots & \vdots & \vdots & $\ddots$ & \vdots \\
      $k_n$ & $k_{n_1}$ & $k_{n_2}$ & $\ldots$ & $k_{n_g}$ \\
      \hline \hline
    \end{tabular}
\end{table}

\item Nodes work among themselves to pool the appropriate covariate
  values according to the case-id and control-id datasets and pass
  that to the analytic center. 

\item Analytical center uses a dataset as in Table \ref{tab:PooledData} and
  runs logistic regression using case pool status (yes/no) as outcome
  and pooled covariate values.
\end{enumerate}
\begin{table}
\caption{Pooled data passed from the nodes to the Analytical Center
  that is to be used for PoLoR.}
\label{tab:PooledData}
\begin{center}
    \begin{tabular}{lcc}
      \hline \hline
      Case Pool & Pool id & Pooled Covariate value \\ \hline
      Yes & $1$ & $s^1_1$ \\
      \vdots  & $\ddots$ & \vdots \\
      Yes & $k_n$ & $s^1_{k_n}$  \\ \hline
      No & $1$ & $s^0_1$ \\
      \vdots  & $\ddots$ & \vdots \\
      No & $k_m$ & $s^0_{k_m}$ \\
      \hline \hline
    \end{tabular}
\end{center}
\end{table}

In addition to a primary exposure, confounders can be accommodated in
a similar fashion as outlined before. In an epidemiological context,
only categorical effect modifiers can be accommodated in
PoLoR. Moreover, transformation of the covariate cannot be handled in
PoLoR.  
In contrast, in the context of patient privacy, these
issues can be addressed because individual level data is already
collected and is available at the node level. It is easy to see that
by using random pooling conditional {\it only on the outcome}, we can 
essentially circumvent these issues to 
(1) estimate all the components of the effect modifiers including the
main effect and interaction and (2) accommodate continuous effect
modifiers. Transformations can be handled in a similar manner: instead
of sending $\sum_i x_i$, pooled level of appropriate transformation,
such as $\sum_i \log(x_i)$ can be made available to the center. Care
should be taken in the choice of $g$ when multiple functional forms of
the same covariates are included in the model. For example, if a cubic
power and all lower powers of exposure $X$ is included in the model,
then use of $g \leq 3$ would divulge the individual patient covariate
values. However, even in this case, it will not be possible to extract
exact covariate combination for an individual participant.

As mentioned earlier, multiple pool sizes can be accommodated with the
only stipulation that at least one pool of a specific size should be
present in both case and control pools. However, unlike the standard
approach, due to privacy concerns, we would not be able to create
pools of size 1 in distributed data setting. We recommend choosing
pool sizes to exhaust both $n$ and $m$; otherwise, excluding a very
small number of observations from cases and controls. Not including
these small number of samples in the analysis will likely minimally
impact estimation and efficiency and will still keep patient records
private. For example, if there are $100$ cases and $4321$ controls,
one can use pool sizes of $3$ and $4$, with $4$ pools of size $3$ and
$22$ pools of size $4$ among cases and $3$ pools of size $3$ and
$1078$ pools of size $4$ among controls. Otherwise, one can use pool
of size $4$ only and exclude one control observation altogether from
consideration.

In a distributed data setting, much of the limitation of the
traditional setting, in particular for model selection, can be
circumvented simply because individual level data can be accessed at
the node level. For example, choosing between alternate
transformations of the primary covariate of interest or confounder and
EM would only require that center resend the new functional form to
the nodes and nodes send back the pooled values according the
transformation, while keeping the pool memberships the same as
before. Hence, all standard diagnostics for logistic regression model
can be adapted here. However, appropriateness of logit link still
cannot be assessed in this setting using the existing tools. 

\noi{\bf Additional Considerations for Privacy}\\
\underline{Secure summation:} In addition to pooling of the covariate
value, secure summation can be employed to further protect individual
records. A simple secure 
summation protocol would require generation of random numbers for each
pool and/or each node (and/or each covariate). After pooling the
covariate level for a given pool, the node adds a random number to the
pooled covariate value and passes this on to the next node. The next
node, adding the covariate values from appropriate pools, again add a
random number and pass on to other node as appropriate. When the
aggregate levels of the covariates reach the center, the nodes also
pass along the set of random number that were added to the
pools. Finally, the center subtracts the random numbers as appropriate
and works with the aggregate covariate values. If multiple nodes are
not colluding, secure summation is a very secure process. For more
details about secure summation, see, \cite{KarrEtAl2007Techno}.
\\\underline{Randomizing the order of the nodes for summation:} For a
given pool, when multiple covariates need to be aggregated, we can
randomly order the nodes to add another layer of security. For
example, if nodes 1, 2, and 3 contribute to pool 1 for two variables,
then for one of the variables, the secure summation may begin with
node 1, then node 3 and finally end with node 2, while for the second
variable, the secure summation can follow the order: 3, 1, 2. Of
course, only the center should have the information about the
randomization scheme while nodes could have a limited information so
that they can pass the aggregate value to the appropriate nodes.

\section{Example}
\label{sec:example}
\subsection{Simulated Data}
\label{subsub:Distributed}

To demonstrate the viability of pooled analysis in a setting where
individual-level covariate information can be accessed but not shared,
we carried out simulation studies. Our earlier 
research and that of our collaborators established that pooled
analysis produces consistent estimator of an exposure subject to
pooling, a confounder (may or may not be subject to pooling) and
interaction parameter for a categorical EM
\citep{WeinbergUmbach1999Biometrics,  
SahaPooledTTP2010, SahaPooledMatching2011}. 
For a binary outcome, we demonstrate the feasibility of pooling in a
distributed data context with transformation of a confounder and show
that all parameters associated with a continuous EM can be
consistently estimated using pooled analysis. 

We considered $500$ simulated datasets each with a sample size of
$30000$. The primary exposure $X$ was assumed to be normally
distributed. The confounder $Z_1$ was distributed as an absolute value
of  standard normal variable that had a correlation of $0.3$ with the
primary exposure. Finally, a continuous EM $Z_2$ was generated as
standard normal independent of both $X$ and $Z_1$. The following model
was considered for the binary outcome generation:
\begin{eqnarray*}
  \mbox{logit}\Pr(Y = 1|x, z_1, z_2) = \beta_0 + \beta_x x + \beta_{z_1}
  log(z_1) + \beta_{z_2} z_2 + \beta_{xz_2} x~ z_2. 
\end{eqnarray*}
We considered several combination of parameter values and show the
results for the following set: $\beta_0 = -3.0, \beta_x = 0.25,
\beta_{z_1} = -0.3, \beta_{z_2} = 0.15, \beta_{xz_2} = 0.5$, resulting
in a prevalence of $6.8\%$. We used all observations for standard
analysis. 

\begin{figure}[!htbp]
  \centering
  \includegraphics[scale=1.0]{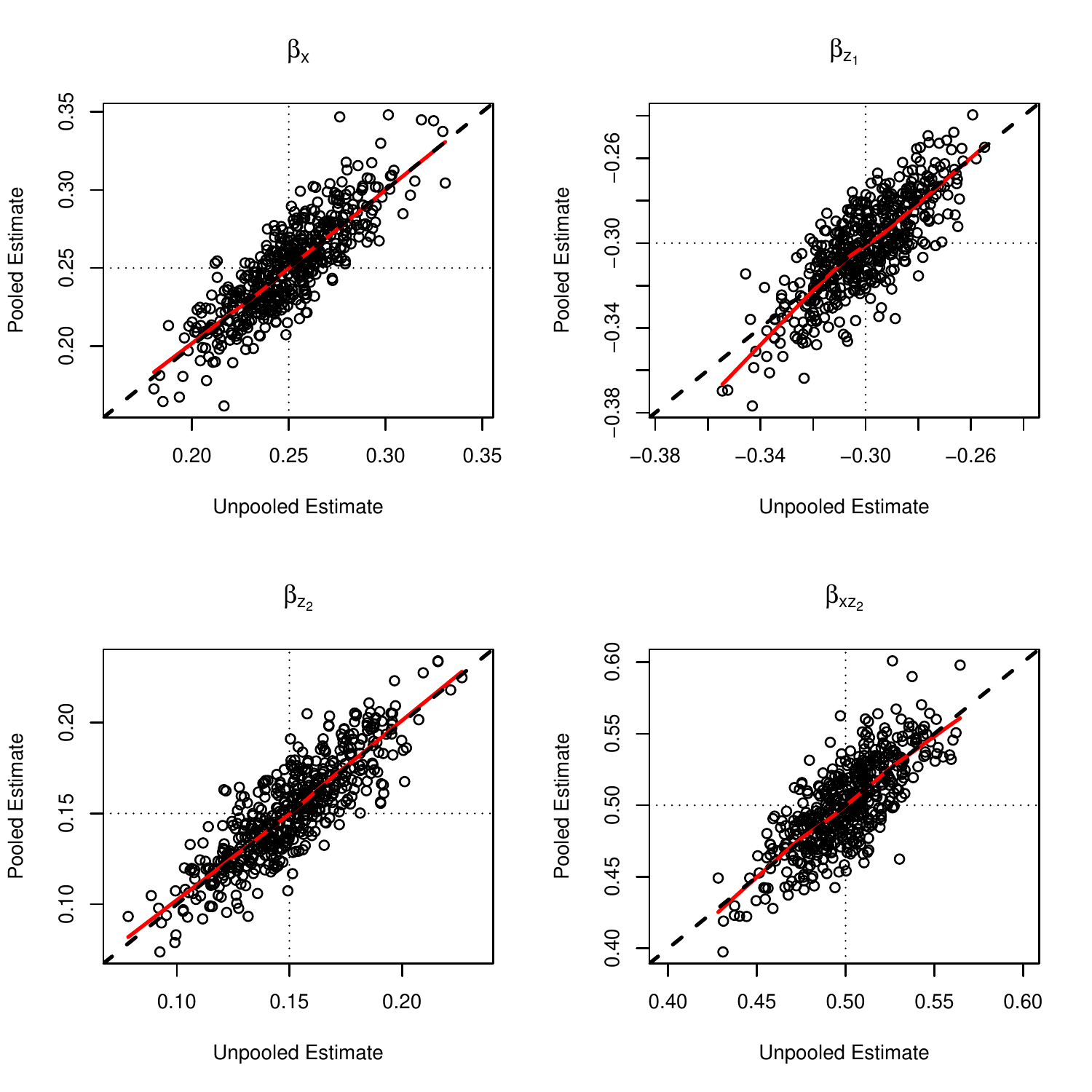}
  \caption{Comparison of individual parameter estimates between
    standard analysis and pooled analysis with $g=6$ for a binary
    outcome for four model parameters (see section
    \ref{subsub:Distributed} for detailed simulation setting). The
    true parameter values on both axes are indicated by dotted
    lines. A lowess line is added in red and the dashed black line has
    a slope of $1$ and intercept of $0$. There is good agreement
    between the black and red lines for all parameters indicating that
    the parameter estimates from pooled analysis are generally
    comparable to that from standard analysis for each of the $500$
    simulations.} 
  \label{fig:DistributedCC}
\end{figure}

For pooled analysis, we considered pools of sizes: $g = 2, 3, 4, 6$
conditional on the outcome status. To simplify the setting, we
discarded at most $(g-1)$ observations from cases and/or controls when
the $n$ and/or $m$ were not divisible by $g$. Simulation for the
entire setup was performed in the programming 
language R. In Table \ref{tab:PoolDistributed} we report the average
parameter estimate, average model-based standard error (SE) and
coverage probability 
out of $500$ simulations. In Figure \ref{fig:DistributedCC}, we
plotted the parameter estimates from the standard analysis and pooled
analysis (with $g = 6$) to assess the agreement between the two
sets of parameter estimates.   

\begin{table}
  \caption{Parameter estimates between standard analysis and pooled
    analysis with different poolsizes for a binary outcome. (see section
    \ref{subsub:Distributed} for detailed simulation setting). In
    addition, the model-based SE (ModelSE) and coverage (nominal:
    $0.95$) are also shown. The power (not shown) for all parameters
    and for both unpooled and pooled analysis are equal to $1$.} 
  \label{tab:PoolDistributed}
  \centering
  \begin{tabular}{lrrrrr}
    \hline \hline 
         &           & \multicolumn{4}{c}{Pooled} \\\cline{3-6}
$\beta_x = 0.25$ &  Unpooled &    g=2 &    g=3 &    g=4 &    g=6\\ \hline
Estimate &   0.2499 & 0.2500 & 0.2500 & 0.2504 & 0.2502    \\
ModelSE  &   0.0245 & 0.0253 & 0.0262 & 0.0272 & 0.0293    \\
Coverage &   0.9540 & 0.9400 & 0.9560 & 0.9400 & 0.9480    \\ 
\hline
$\beta_{z_1} = -0.30$\\ \hline 
 Estimate & -0.3004 & -0.3007 & -0.3009 & -0.3013 & -0.3022 \\
 ModelSE  &  0.0175 &  0.0184 &  0.0193 &  0.0203 &  0.0224 \\ 
 Coverage &  0.9580 &  0.9580 &  0.9520 &  0.9540 &  0.9520 \\
\hline
$\beta_{z_2} = 0.15$\\ \hline 
 Estimate &  0.1507 & 0.1508 & 0.1513 & 0.1504 & 0.1514  \\
 ModelSE  &  0.0243 & 0.0251 & 0.0259 & 0.0268 & 0.0288  \\ 
 Coverage &  0.9620 & 0.9540 & 0.9500 & 0.9520 & 0.9700  \\
\hline
$\beta_{xz_1} = 0.50$\\ \hline 
 Estimate &  0.5002 & 0.5006 & 0.5003 & 0.5016 & 0.5005\\
 ModelSE  &  0.0225 & 0.0237 & 0.0250 & 0.0264 & 0.0294\\ 
 Coverage    &  0.9320 & 0.9340 & 0.9360 & 0.9220 & 0.9440
\\ \hline \hline
  \end{tabular}
\end{table}

Comparing the parameter estimates, model-based SE and
coverage for the four parameters of the model between standard
logistic regression and PoLoR, we see that on an average PoLoR
performs comparably to standard analysis. Given that pooling
essentially reduce  
the data dimension, we see a tendency of increased model-based SE,
especially for larger poolsizes. However, the corresponding coverage
of the model-based CI is generally close to the nominal coverage.

We also plotted the estimates of the four parameters ($\beta_x,
\beta_{z_1}, \beta_{z_2}, \beta_{x~z_2}$) from each of the
simulations obtained via standard logistic regression and PoLoR with
$g=6$ in Figure \ref{fig:DistributedCC} to 
demonstrate the comparability between two sets of estimates over each
simulation. We added a diagonal line (black, dashed) and the lowess
line (red, solid) to the plots. The red lowess line is generally in
agreement with the diagonal black line indicating the similarity
between the two sets of estimates.

\vspace{5mm}
\noi {\bf Practical Consideration: Poolsize Recommendation}\\
In the simulation settings we considered with a sample size of 30,000,
poolsizes $g$ between $10-40$ produced  PoLoR estimates that were
comparable to the estimates from standard logistic regression. Due to
regression to the mean, larger poolsizes, especially, for a skewed
covariate, produced estimates that were biased away from
null. Therefore, we do not recommend poolsizes larger than
$40$. Poolsizes $g=10-20$ (or smaller) performed well for different
exposure distributions. For distributed data setting, $g$ between
$5-20$ could be used to achieve a balance between privacy and unbiased
estimation. 

\subsection{Real Data}
We demonstrate the comparability of PoLoR with standard logistic
regression using a modified version of data from one of the first
successful trials of adjuvant therapy regimens for colon cancer.  The
original study compared two treatment: Levamisole, and combination of
Levamisole and Fluorouracil against observation only arm using a
randomized controlled clinical trial among stage III colon cancer
patients \citep{Colon1989}. Due to strong evidence of benefit from
the combination of Levamisole and Fluorouracil, the trial was stopped
early. The dataset ({\tt data(colon)}) has been downloaded from R
library {\tt survival} and is closest to that of the final report by
\cite{Colon1995}. 

In the original dataset, each patient contributed
two time-to-event outcomes: time to recurrence and time to
survival. We use only the time to recurrence and converted the
endpoint to a binary outcome using ``recurrence-free for 5 years''
(yes/no) as an endpoint. Fifty five patients censored before 5 years
and 22 patients with missing tumor 
differentiation status were excluded, leaving 866 patients. For
standard logistic regression and PoLoR, we included the following
variables in 
the model: sex (female:0, male:1), age (in years), obstruction of
colon by tumor (yes/no), perforation of colon (yes/no), adherence of
tumor to nearby organs (yes/no), tumor differentiation (1=well,
2=moderate, 3=poor), more than 4 positive nodes (yes/no), and treatment
(observation only, Levamisole and Levamisole+Fluorouracil).

For PoLoR, we aggregated the variables using poolsizes $g=3$ and $g=4$
resulting in $288$ and $216$ pools respectively. Two patients were
excluded at random to create the pools. Larger poolsizes were not
considered due to relatively modest sample size of 866.  The pools
were created randomly. The parameter estimates of PoLoR are dependent on
the particular pool formation and can differ according to different
pool formation. We report the estimates based on one particular
realization. For demonstration
purpose, we considered age as a confidential variable. We created pool
ids at random and aggregated all variables except age according to the 
pool ids. Instead of aggregating age, we aggregated a perturbed age
variable by adding a random integer between 1 and 20 to the observed
age. Aggregate age was then derived from the perturbed age by
subtracting the aggregate of the random numbers. Finally we analysed
the pooled data using PoLoR with the appropriate offset. We list the
log OR and the SE for each of the covariate for standard
logistic regression and PoLoR ($g=3, 4$) in Table \ref{tab:Colon}.
In general, the direction and magnitude of the log ORs are similar between
standard logistic regression and PoLoR, except for the variables
perforation (only 27 patients with perforation) and differentiation (91
patients without differentiation). The SEs are generally comparable. 
  
\begin{table}[htbp!]
  \caption{Log OR and SE for standard logistic regression and PoLoR for Colon cancer data.}
  \label{tab:Colon}
  \centering
  \begin{tabular}[htbp!]{lrrr}
\hline \hline 
& \multicolumn{3}{c}{Log Odds Ratio} \\
& \multicolumn{3}{c}{Standard Error}\\
Variable             & Std logit  & PoLoR $g=3$ & PoLoR $g=4$ \\
\hline
Sex                     & -0.149     & -0.044     & -0.158  \\
                      &  0.144     &  0.163     &  0.180 \\  \hline
Age                      & -0.003     & -0.006     & -0.012  \\
                       &  0.006     &  0.007     &  0.008  \\ \hline
Obstruction                 &  0.096     &  0.117     &  0.143  \\
                  &  0.183     &  0.190     &  0.201  \\ \hline
Perforation                   &  0.466     &  0.531     &  0.154  \\
                    &  0.430     &  0.465     &  0.512  \\ \hline
Adherence                   &  0.408     &  0.346     &  0.612  \\
                    &  0.208     &  0.221     &  0.247  \\ \hline
Differentiation=2       & -0.092     &  0.228     &  0.208  \\
        &  0.236     &  0.265     &  0.272  \\ \hline
Differentiation=3       &  0.163     &  0.632     &  0.603  \\
        &  0.288     &  0.334     &  0.338  \\ \hline
Node4                    &  1.238     &  1.067     &  1.231  \\
                     &  0.171     &  0.187     &  0.215  \\ \hline
Levamisole                    & -0.145     & -0.285     & -0.182  \\
                     &  0.174     &  0.209     &  0.225  \\ \hline
Lev+FU                & -0.750     & -0.788     & -0.880  \\
                 &  0.178     &  0.210     &  0.236  \\
\hline \hline
  \end{tabular}
\end{table}

\section{Discussion}
\label{sec:discussion}
In this manuscript, we introduce PoLoR (\underline{Po}oled
\underline{Lo}gistic \underline{R}egression) as a privacy-preserving
approach to estimate coefficients of a logistic regression model using
only aggregate covariate level. 
PoLoR is an existing idea from observational and epidemiologic study
setting and can be adapted for distributed data network  while
maintaining patient privacy. PoLoR, being based on aggregate covariate
level also reduces data dimension and hence can be used to analyze big
data under a resource-constrained setting.

PoLoR is primarily intended for an exposure that is measured in an assay
(e.g., blood, serum, urine, etc.) and pooling specimen from multiple
study subjects leads to a reduced overall assay cost. 
We show here that covariate pooling approach or microaggregation for
logistic regression 
can be applied in a non-assay setting to protect patient
privacy. PoLoR is an application of specimen pooling for 
distributed data. In PoLoR,
estimation of model parameters is based only on the aggregate
covariate level rather than individual covariate level. Adopting this
strategy in a distributed data setting 
would imply that only aggregate covariate information is needed to
estimate ORs associated with covariates. Since aggregate covariate
levels does not reveal individual information, it can be shared
without compromising patient privacy and hence is suitable in a
distributed data setting. For a binary outcome and a discrete
time-to-event outcome,  the model parameters between
PoLoR and a standard logistic regression based on individual covariate
levels remain the same, thus allowing estimation of the parameters of
interest using aggregate covariate values. 


However, PoLoR is not without limitations. PoLoR is scalable for
larger number of nodes. But to ensure valid parameter estimates, it is
important to keep track of the pools and which patients (from which
node) are contributing to a given pool. For maintaining patient 
privacy in PoLoR, we need to create reasonably large pool size and
therein lies the dilemma of choosing an appropriate poolsize $g$. For
PoLoR, the unit of measurement is pools and not individuals, so the
asymptotic properties of the estimators will be somewhat dependent on
the number of pools rather than the number of individuals. Thus, for
asymptotic properties of the estimators, such as nominal confidence level,
to hold, we need large number of pools, which in turn implies that we
cannot use $g$ that is too large compared to $n$ and $m$, because
$n_k$ and $m_k$ will need to 
be large for asymptotic properties to hold. In our simulations, a
poolsize between $5-20$ achieves a reasonable balance between
unbiasedness and privacy and that is what we recommend. We do not
recommend $g>40$ even for a very large dataset due to regression to mean
and resulting bias in parameter estimates.
If privacy is of no
concern, another important issue would be to assess the properties of
PoLoR when we use different $g$ for cases and control and in
particular use a small value $g_n=1$ of $g$ for cases and a large value
of $g_m > > 1$ for controls with only one pool of size $g_n$ among
controls and one pool of size $g_m$ among cases. Also, PoLoR is meant
for one time application only. Addition of a new patient record would
require creating pools afresh and similar to standard logistic
regression, would require reanalysis. 

Several important issues warrant additional research. One can show that
PoLoR  can be used for a discrete survival outcome. Frequently
though, survival outcome is continuous and currently there is no
method for analyzing survival outcome in a distributed data
setting. Another important extension could be for longitudinal
outcome. Conceivably, PoLoR would likely be applicable in longitudinal
setting, but the details need to be worked out. We surmise that
other approaches in the infectious disease setting that uses pooling
could be adapted for distributed data setting. This may open up
several opportunities for analyzing distributed data. Finally,
approaches from statistical disclosure limitation may be borrowed for
distributed data while maintaining patient privacy.

\section*{Acknowledgement}

\nocite{Olsen2014IntJEpiEditorialDataProtectionEpiRsch,
ElEmamEtA2013JAMIASecureLogistic,
WangEtAl2013JBiomInformaticsEXPLORER,
Wolfson2010IJEDatashield,
Wu2012JAMIAGLORE}

\bibliographystyle{apalike}
\bibliography{/Users/sahap/refs}

\end{document}